\documentclass[reprint, twocolumn, twoside,  nofootinbib, hidelinks, colorlinks, linkcolor=blue, citecolor=blue, urlcolor=black, amsmath,amssymb, floatfix, prl]{revtex4-1}

\usepackage[switch,columnwise]{lineno}
\usepackage{fancyhdr}
\usepackage{helvet}
\usepackage{graphicx}% Include figure files
\usepackage{dcolumn}% Align table columns on decimal point
\usepackage{bm}% bold math
\usepackage{hyperref}% add hypertext capabilities
\usepackage{mathrsfs}
\usepackage{amsbsy}
\usepackage{amsmath}
\usepackage{scrextend}

\begin{document}

\title{Shape Dynamics of Freely Falling Droplets} 
\author{David V. Svintradze}\thanks{Present address: Max Planck Institute for the Physics of Complex Systems, Noethnitzer Str. 38, 01187 Dresden, Germany}
\email{dsvintra@yahoo.com; dvs@pks.mpg.de}
\affiliation{%Max Planck Institute For the Physics of Complex Systems, Noethnitzer Str. 38, 01187 Dresden, Germany 
Department of Physics, Tbilisi State University, Chavchavadze Ave. 03, Tbilisi 0179, Georgia
}
\date{\today}

\begin{abstract}
Oscillating shape motion of a freely falling water droplet has long fascinated and inspired scientists. We propose dynamic non-linear equations for closed, two dimensional surfaces in gravity and apply it to analyze shape dynamics of freely falling water drops. The analytic solutions qualitatively well explain why drops oscillate among prolate/oblate morphologies and display a number of features consistent with experiments. %In addition, dynamic thickening dictates speculative, contradictory suggestion, that cells obtained membranes as byproduct of such surface fluctuations. 
\end{abstract}

\maketitle
%\tableofcontents

Droplets are very peculiar systems mostly due to their shape dynamic properties \cite{deGennes2004}. They conserve their shape at rest due to surface tension,  but are very deformable. Small perturbation of an equilibrium shape may induce large deformations in surface morphologies. One can trigger droplet inertial motion as well as morphological dynamics by inducing substrate surface oscillation \cite{raufaste_2017}. Gradient of free substrate surface energy or angle difference between droplets leading and trailing edges provokes selfpropulsion \cite{Chaudhury1539, PhysRevLett.94.068301}. Droplets dynamical properties and their vibration modes may control their motion over substrate surface \cite{doi:10.1021/la046886s, PhysRevLett.102.194504} and transiently defy gravity \cite{PhysRevLett.99.144501}. The contact with the substrate can be minimized by the development of superhydrophobic surfaces \cite{Vollmer2014}, allowing droplets to bounce on the surface like elastic balls \cite{richard2002}.  

%selfpropulsion is thus observed when the substrate surface exhibits a gradient of free surface energy [2] or when a droplet itself triggers a contact angle difference between its leading and trailing edges [3,4]. 

%Their dynamical properties are at the origin of peculiar behaviors [5] and their vibration modes [6] can be forced in a way to control their motion over a substrate [7,8] and to let them move against gravity [9]. 

%The contact with the substrate can also be prevented by triggering the Leidenfrost effect [10,11], which leads to nonsticky drops that move easily [12]. In the same vein, the development of superhydrophobic surfaces (SHSs) allows one to minimize the adhesion with the substrate [13] and droplets can bounce on such surfaces like elastic balls with a velocity restitution coefficient that depends on the relative importance of inertial and capillary effects [14]. Recently Boreyko and Chen [15] have studied the coalescence of droplets during vapor condensation on SHSs. The surface energy release associated with the coalescence is partly transformed into kinetic energy and induces a vertical propulsion of the merged drops.

%Shape dynamics of freely falling water drops (sometimes referred as wobbling motion)
%Wobbling motion of bubbles, water drops dancing on the bath surface, soap films fluctuating while flying in an air, 
%have long been fascinating subjects to study. 

In addition, recent experiments demonstrate that an oscillating surface can launch wobbly water drop into the air at higher speed than it would launch a hard ball of the same mass \cite{raufaste_2017}. Therefore, a synchronization between the internal vibration of a drop projectile and the frequency of the rising and falling surface can more than double kinetic energy of the droplets. 
 
The behavior of bubbles differs from drops in gas-liquid systems. They display much broad range of shape deformations including turbulence \cite{kellay_2002}, thickness variations \cite{Greffier_2002,rivera_1998,nierop_2008}, the Marangoni effect \cite{tran_2009}, draining \cite{moulten_2010}, ejection of droplets \cite{drencken_2008}, rupture \cite{debr_1995}, self-adaptation \cite{boud_1999}, chaotic behavior \cite{gilet_2009}. From these broad range of effects thickness variations have been explained by numerical solutions of dynamic nonlinear equations for free thin fluid films \cite{grinfeld2010}.

Despite of numerous experimental data \cite {deGennes2004, raufaste_2017, Chaudhury1539, PhysRevLett.94.068301, doi:10.1021/la046886s, PhysRevLett.102.194504, PhysRevLett.99.144501, Vollmer2014, richard2002}, description of droplets shape dynamics remain analytically largely unsolved. Some dynamic effects in linear regimes have led to classical wave equations \cite{isenberg1992, chomaz1998}. Numerical simulations of Navier-Stokes proved to be effective tool in the path of describing shape motions of drops and bubbles \cite{tripathi2015, agrawal_2017, tripathi2014}. Furthermore, numerical solution of one dimensional Navier-Stokes equations showed that simulated shape dynamics agrees with experiment qualitatively \cite{eggers1994drop}. Also, combination of experiments and numerical simulations explained the shapes of singularities around the neck for a drop falling from a faucet \cite{Shi219}.

In this letter we introduce a formalism to describe closed, two dimensional surface dynamics and analytically address morphological patterns of droplets observed in experiments \cite {deGennes2004, raufaste_2017, Chaudhury1539, PhysRevLett.94.068301, doi:10.1021/la046886s, PhysRevLett.102.194504, PhysRevLett.99.144501, Vollmer2014, richard2002}. Our model predicts transient excursions against the gravitational force
 %Also we explain why gravity has such little or no role in shape dynamic effects for droplets 
 and indicate why they can move against gravity. We limit ourself with shape dynamics of droplets, though due to generality of our arguments analyses can be extended to wobbly dynamics of bubbles too.

 %For a sake of simplicity, uniform translational motion is ignored in our equations, but an analytic solution to shape dynamics does demonstrate  

%A blob of fluid, moving in the medium, is called bubble if it's density is less than its surroundings and it is called drop otherwise. The behavior of bubbles differs from drops in gas-liquid systems. Initially spherical air or soap film bubble rising in liquid or air can either spiral or zigzag \cite{saffman_1956, zenit_2008, cano_2016}, and can display arrays of static and dynamic physical effects including turbulence \cite{kellay_2002}, thickness variations \cite{Greffier_2002,rivera_1998,nierop_2008}, the Marangoni effect \cite{tran_2009}, draining \cite{moulten_2010}, ejection of droplets \cite{drencken_2008}, rupture \cite{debr_1995}, self-adaptation \cite{boud_1999}, chaotic behavior \cite{gilet_2009}. Thickness variations have been explained by numerical solutions of dynamic nonlinear equations for free thin fluid films \cite{grinfeld2010}. In contrast to bubble, initially spherical liquid drop in air falls in a straight path \cite{vries2001}.

%We propose dynamic equations for closed, two dimensional surface motions in gravity and by analytical solutions demonstrate same morphological pattern as observed in \cite{raufaste_2017} (see also \cite{richard2002}), and provide avenue of describing this interesting experiment. 

We address shape dynamics of freely falling water droplets analytically. Droplet is bouncing vertically from super hydrophobic substrate. The substrate itself is allowed to move freely. For this problem, instead of solving Navier-Stokes equations numerically, which have already been done \cite{tripathi2015, agrawal_2017, tripathi2014}, we use dynamic nonlinear equations for moving two dimensional surfaces \cite{svintradze_2017, svintradze_2016} and solve it analytically. We only concentrate on the shape dynamics and ignore path instabilities, by setting interactions with the environment negligibly small. Of course, this set up might be comparable to liquid droplet falling in an air, but not to dynamics of an initially spherical bubble rising in quiescent liquid, where one can not neglect interactions with environment. We consider a water drop as a continuum medium with single uniform surface and, in order to simplify analyses neglect internal friction, which by experiments proved to have insignificant effect \cite{raufaste_2017}. 
% which would be incorporated into equations as interaction between two or many surfaces and could be modeled as visco-elastic effects.   

 The formalism which ultimately leads to derivation of differentially variational surface (DVS) equations \cite{svintradze_2017, svintradze_2016}, for freely falling droplets, generalizes Eulerian representation of fluid dynamics and in contrast to Navier-Stokes, as it is demonstrated in this paper, is analytically solvable. The formalism is fully covariant and its analytical solutions qualitatively exactly reproduces the surface motion. 
 
\begin{figure}
	\includegraphics[scale=1.5]{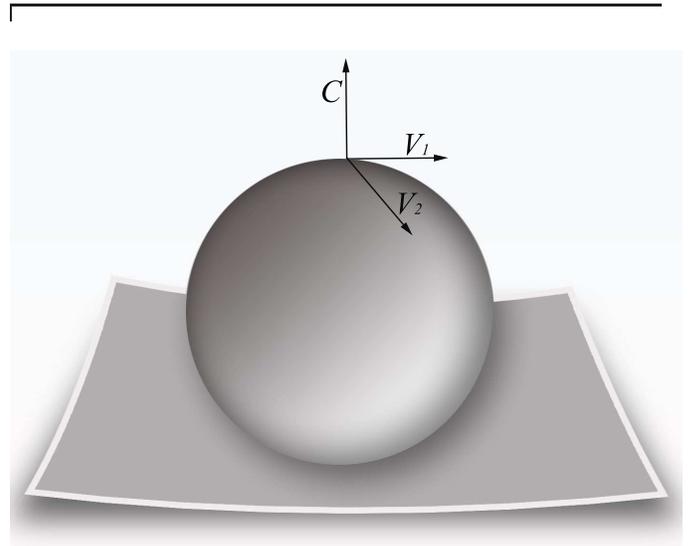}
	\caption{\label{fig:1} Graphical illustration of a droplet sitting on a superhydrophobic substrate. Equilibrium shape of a droplet in a rest is a sphere. Surface velocities $C, V_1, V_2$ for a arbitrary chosen point are shown by arrows.}
\end{figure}

 As far as the body falls freely in the gravity, we need to add gravitation to the equations of two dimensional surface motion and therefore, the differential variation of surface (DVS) equation \cite{svintradze_2017, svintradze_2016} (see also \cite{Svintradze_2018}) should be modified accordingly. The Lagrangian of the motion reads:
\begin{equation}
L=\int_S \frac{\rho_S V^2}{2}dS+\int_\Omega (p+\rho gh)d\Omega \label{1}
\end{equation}
where $\rho_S$ is a surface mass density, $V$ is a surface velocity, $p$ pressure across the surface, $\rho gh$ is a hydrodynamic pressure applied by gravity and $\rho$ is water drop mass density. The potential energy term is modeled as negative volume integral from the pressure. In other words, relation between surface pressure and potential energy is defined as

\begin{equation}
U=-\int_\Omega pd\Omega \nonumber
\end{equation}
Variation of the Lagrangian (\ref{1}) modifies the DVS equations \cite{svintradze_2017, svintradze_2016} so that the gravitation is taken into account.  As far as DVS equations are already derived \cite{svintradze_2017, svintradze_2016}, instead of taking brute mathematical steps we just mention how gravitational term $\rho g h$ modifies  final equations, which reads: 
\begin{widetext}
\begin{align}
\dot{\nabla}\rho_S+\nabla_i(\rho_S V^i)&=\rho_S CB_i^i \nonumber \\
\partial_\alpha(V^\alpha(\rho_S (\dot{\nabla}C+2V^i\nabla_iC+V^iV^JB_{ij})+p+\rho gh))&=-V^\alpha\partial_\alpha(p+\rho gh) \label{2} \\
\rho_S V_i (\dot{\nabla}V^i+V^j\nabla_jV^i-C\nabla^i C-CV^jB_j^i)&=-N^\alpha V_i\nabla^ip_\alpha \nonumber
\end{align}
\end{widetext}
where $p_\alpha=(p+\rho gh)N_\alpha, \alpha=1,2,3$. The first equation is a generalization of the conservation of mass, the second and third equations display motion in normal and tangent directions of the closed surface, and all satisfy conservation of mass and energy. $\dot{\nabla}=\partial/\partial t-V^i\nabla_i, i=1,2$ is curvilinear, invariant time derivative, $\nabla_i$ is curvilinear derivative, $C$ is surface normal velocity, $V_i$ is component of surface tangent velocity, $V^\alpha$ stands for ambient component for surface velocity, $N^\alpha$ is component of unit surface normal, $B_{ij}$ is curvature tensor and $B_i^i$ is mean curvature. Surface velocities are illustrated on Fig.~\ref{fig:1}. Repeated indexes indicate Einstein summation convention. The equations (\ref{2}) are covariant and are valid not only for capillary surfaces, but for moving surfaces of molecules too \cite{svintradze_2017}. 

Note that (\ref{2}) violates Newton laws. As an example we shall demonstrate violation of first law, informally stating: no force no/or constant velocity. Translating this statement for the surfaces one should expect, that if there is no potential field acting on the surface, than the surface should be either in a rest or surface velocity has to be constant. In contrast, what we find is that if one removes potential fields, i.e. sets $\rho gh$ and $p$ naught in equations, then one should expect that a solution to the equations must be constant normal $C$ and constant tangent $V_i$ velocities. It is easy to check that the constant surface velocity indeed satisfies the equations of motions for freely floating droplets, but they are not the only solutions. Freely floating water drops will continue shape dynamics (if they were moving \textit{a priori}) and retain non-linearity even in the case when no potential field (a force) acts on them.

Note that the gravity acts only in normal direction, so that tangential gradient of the surface pressure can be modeled as negligibly small:
\begin{equation}
-V_iN^\alpha\nabla^ip_\alpha=0 \label{3}
\end{equation}
As we have already shown in our previous works \cite{svintradze_2017, svintradze_2016}, the second equation significantly simplifies if the surface is homogeneous and can be described with time invariable surface tension $\sigma$, then 
\begin{equation}
\rho_S(\dot{\nabla}C+2V^i\nabla_iC+V^iV^jB_{ij})=\sigma B_i^i \label{4}
\end{equation}
We refer to (\ref{4}) as dynamic fluid film equation for surface normal motion \cite{grinfeld2010, grinfeld2013}. Using (\ref{4}) in (\ref{2}), we end up with 
\begin{equation}
\partial_\alpha[ V^\alpha(\sigma B_i^i+p+\rho gh)]=-V^\alpha\partial_\alpha (p+\rho gh) \label{5}
\end{equation}
Eq. (\ref{5}) is a solution for the second equation indicating the surface motion in normal direction. We now assume that deformations along tangent directions are negligibly small compared to normal ones, then equations (\ref{2}) with condition (\ref{3}) simplify as:
\begin{align}
\frac{\partial\rho_S}{\partial t}&=\rho_SCB_i^i \label{6} \\
\partial_\alpha[V^\alpha (\sigma B_i^i+p+\rho gh)]&=-V^\alpha\partial_\alpha (p+\rho gh) \label{7} \\
\rho_S\frac{\partial C}{\partial t}&=\sigma B_i^i \label{normal}
\end{align}  
%\begin{figure}
%\includegraphics[scale=6]{figure_1}
%\caption{\label{fig:1} Photograph of freely falling water drop on bath surface. When the drop touches the surface for a short period of time it adopts an equilibrium shape a sphere. After spending some energy on shape deformations and some energy dissipates in the bath and in internal energy of the drop, the drop jumps again. After the jump the drop is driven out of equilibrium shape and will start shape changing motion again.}
%\end{figure}
Where (\ref{normal}) comes from (\ref{4}) with assumption that tangent velocities are infinity small. In this case (\ref{7}) is the solution of (\ref{6},\ref{normal}). 

When the droplet touches the ground (substrate surface), for very short period of time, the gravity becomes compensated by a substrate surface reaction force, so that, the drop comes in equilibrium with the substrate and reaches stationary shape 
%(Fig.~\ref{fig:1}), 
satisfying conditions:  $C=0, \partial_\alpha V^\alpha=0$ and $\partial (p+\rho gh)/ \partial t=0$, then solutions to (\ref{6},\ref{7}) are:  
\begin{align}
\rho_S&=\rho_0=const \label{8} \\
B_i^i&=-\frac{p+\rho gh}{\sigma} \label{9}
\end{align}
Equation (\ref{8}) dictates that water molecules are homogeneously distributed on the surface, and (\ref{9}) shows that the shape adopts constant mean curvature. Water droplets are closed and compact surface, therefore by the Alexandrov theorem \cite{alexandrov1958} the shape with constant mean curvature must be a sphere. Needless to say that droplet sitting on a superhydrophobic substrate is indeed perfect sphere \cite{deGennes2004}. Note that by solving (\ref{7}) analytically we mathematically proved that it must be sphere, while Young-Laplace law can not be considered as proof.  

 %and it is according to experiments. %(Fig.~\ref{fig:1}) 

Before we proceed further, note that (\ref{6}-\ref{8}) are consistent with existing infinitesimal models \cite{isenberg1992, chomaz1998}. Indeed, let $\rho_0$ be equilibrium surface density and assume that both $C$ and $B_i^i$ are infinitely small, so that linearized conservation of mass (\ref{6}) reads $\partial\rho_S/\partial t=0$ with solution $\rho_S=\rho_0$. This indicates that for small enough $C$, the density of the diffusive layer remains constant and infinitesimal models, in the limit of small oscillations and small mean curvature, are consistent with our framework.

Furthermore, if the mean curvature $B_i^i$ is time dependent functional, then according to calculus of moving surfaces
\begin{equation}
	\frac{\partial B_i^i}{\partial t}=\Delta C+CB_{ij}B^{ij} \label{curvature derivative}
\end{equation}
where $\Delta=\nabla_i\nabla^i$ is surface Laplacian \cite{grinfeld2013}. For surfaces with vanishing mean curvature, $B_{ij}B^{ij}$ becomes negative twice the Gaussian curvature $K$ \cite{grinfeld2010}. Therefore, by differentiating (\ref{normal}) linearized acceleration becomes
\begin{equation}
\frac{\partial^2C}{\partial t^2}=\gamma(\Delta C-2KC) \nonumber
\end{equation}  
where $\gamma=\sigma/\rho_0$. In the limit of planar cut off, i.e for flat equilibrium configurations, the Gaussian curvature is zero $K=0$, and infinitesimal deformations are governed by the wave equation:
\begin{equation}
	\frac{\partial^2 C}{\partial t^2}=\gamma\Delta C \label{linear}
\end{equation}
 With the sign convention equation (\ref{linear}) is exactly a case for linear models and is consistent with \cite{isenberg1992, chomaz1998}.

After the droplet spends some energy on surface deformation, it jumps again, but since it has already found equilibrium, which is sphere, it will start oscillation around it like a pendulum. Since, the shape oscillates near to equilibrium, we can suggest that according to (\ref{9}) the term $\sigma B_i^i+p+\rho gh$ becomes infinitely small constant, therefore (\ref{7}) can be modified as
\begin{align}
\partial_\alpha[V^\alpha (\sigma B_i^i+p+\rho gh)]&=-V^\alpha\partial_\alpha(p+\rho gh) \nonumber \\
&=-(\partial_\alpha[V^\alpha(p+\rho gh)] \nonumber \\
&-(p+\rho gh)\partial_\alpha V^\alpha) \nonumber \\
\partial_\alpha [V^\alpha(\sigma B_i^i +2(p+\rho gh))]&=(p+\rho gh)\partial_\alpha V^\alpha \label{10}
\end{align}
Since $\sigma B_i^i$ can not be zero and its combination with surface and hydrodynamic pressure is quasi infinitely small constant, then the solution to (\ref{10}) is 
\begin{equation}
\partial_\alpha V^\alpha=0 \label{11}
\end{equation}
This solution is the one which is consistently identified as continuity condition in Navier-Stokes equations \cite{LIFSHITZ1987ix}. However, setting $\sigma B_i^i +2(p+\rho gh)$ as infinitely small constant is an approximation and therefore, (\ref{11}) is not an exact solution.  (\ref{11}) shows that zero divergence of surface velocity is the near equilibrium solution of DVS for water drops under gravity. (\ref{11}) can be trivially handled by $\bm{V} = k\bm{R}/R^3$ functional (where $k$ is some constant and $\bm{R}$ is position vector), which corresponds to sphere in equilibrium stationary case. Therefore, in ($0\leq\theta\leq2\pi, 0\leq\phi\leq\pi$) spherical coordinates, assuming designation $\bm{S}=(\sin\phi \sin\theta, \sin\phi \cos\theta, \cos\theta)$ the solution reads:
\begin{equation}
\frac{\partial \bm{R}}{\partial t}=\omega_\xi R_\xi\bm{S^\xi} \label{12}
\end{equation} 
where $\omega_\xi, \xi=x,y,z$ are some frequency like functionals and for the sake of simplicity we consider them as constants. $\bm{S^\xi}$ is base vector of $\bm{S}$ and $R_\xi$ is projection of position vector in $\xi$ direction. Trivially, (\ref{11}, \ref{12}) leads to solutions: for some $R_\xi$ constants  with initial condition ${\bm{R_{t=0}}=\bm{R_0}}=R_0\bm{S}$ and  $R_\xi=R_\xi(t)$ time variable functionals, reading respectively:
\begin{align}
\bm{R}=&\bm{R_0}+\omega_\xi R_\xi t\bm{S^\xi} \label{13} \\
%(\omega_xR_xt\sin\phi \sin\theta, \omega_yR_yt\sin\phi \cos\theta, \omega_zR_zt\cos\theta) \nonumber \\
%+R_0(\sin\phi \sin\theta, \sin\phi \cos\theta, \cos\theta) 
\bm{R}=&R_{0\xi}e^{\omega_\xi t\bm{S^\xi}}  \label{14}
%(R_{0x}e^{\omega_xt\sin\phi \sin\theta}, R_{0y}e^{\omega_yt\sin\phi \cos\theta}, R_{0z}e^{\omega_zt\cos\theta}) 
\end{align}
where $e^{\omega_\xi t\bm{S^\xi}}$ is defined as the vector components of which are exponents of frequency times time and spherical coordinate: $e^{\omega_\xi t\bm{S^\xi}}=(e^{\omega_xt\sin\phi \sin\theta}, e^{\omega_yt\sin\phi \cos\theta}, e^{\omega_zt\cos\theta})$. 
\begin{figure}
\includegraphics[scale=1.5]{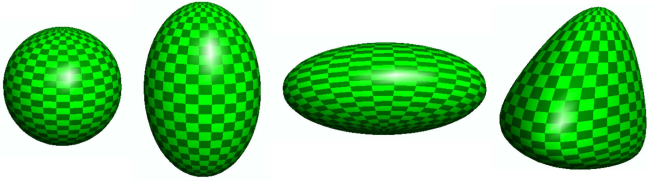}
\caption{\label{fig:2} Illustration of analytically solved shapes. From left to right: sphere with unit radius ${\bm{R_0}=1}$ (initial condition of (\ref{13}) solution), prolate shape drop $\bm{R}=(\sin\phi \sin\theta, 0.86 \sin\phi \cos\theta, 1.75\cos\theta)$ (solution (\ref{13}) at some given time), oblate shape $\bm{R}=(\sin\phi \sin\theta, 1.72 \sin\phi \cos\theta, 0.75\cos\theta)$ (solution (\ref{13}) at other moment), non-prolate/oblate shape $\bm{R}=(e^{\sin\phi \sin\theta}, e^{\sin\phi \cos\theta}, e^{\cos\theta})$ (solution (\ref{14}) at some given time).}
\end{figure}
(Fig.~\ref{fig:2}) is a graphical representation of (\ref{13},\ref{14}) solutions. 

In addition to (\ref{13},\ref{14}), according to vector calculus analytic solution to (\ref{11})  is
\begin{equation}
\bm{V}=-\bm{\nabla}\times\bm{\mathcal{L}} \label{15}
\end{equation}
where $\bm{\mathcal{L}}$ has same dimension as angular momentum. Suggesting that $\bm{\mathcal{L}}$ should behave same way as $\bm{R}$, with conserving full generality, the solution (\ref{15}) can be rewritten as
\begin{align}
\frac{\partial\bm{R}}{\partial t}&=-\bm{\nabla}\times\bm{\mathcal{L}} \nonumber \\
\frac{1}{v_0^2}\frac{\partial\bm{\mathcal{L}}}{\partial t}&=\bm{\nabla}\times\bm{R} \label{16}
\end{align}  
where $v_0$ is some constant wave propagation velocity. After taking the curl of (\ref{16}) and using the curl of the curl ($\bm{\nabla}\times\bm{\nabla} $) identity\footnote{According to curl of curl identity $\nabla\times(\nabla\times\cdot)=\nabla(\nabla\cdot)-\nabla^2\cdot$, applying this to (\ref{16}) one obtains $\nabla(\nabla\bm{R})-\nabla^2\bm{R}=-1/v_0^2\partial^2\bm{R}/\partial t^2$. Taking into account (\ref{11}) $\nabla\bm{R}=const$, therefore first term vanishes and one ends up with (\ref{17}).} one ends up with a wave equation for position vector:
\begin{equation}
\frac{1}{v_0^2}\frac{\partial^2\bm{R}}{\partial t^2}-\nabla^2\bm{R}=0 \label{17}
\end{equation}
Taking into account that as the time evolves all spatial directions become linearly independent $R_\xi=\psi (\xi,t)$ (where $\xi=x,y,z$ and $\psi$ is some functional),  then (\ref{17}) transforms as one dimensional wave equation $1/v_0^2\partial^2\psi/\partial t^2=\partial^2\psi/\partial\xi^2$  for $\xi=x,y,z$. Applying boundary $\psi(0,t)=0, \psi(L,0)=0$ and initial conditions $\psi(\xi,0)=f(\xi), \partial\psi/\partial t(\xi,0)=g(\xi)$ one obtains a solution
\begin{align}
\psi(\xi,t)=\sum_{m=1}^{\infty}\frac{2}{L}\left(\int_0^L\psi(\xi,0)\sin\frac{m\pi \xi}{L}d\xi\right) \nonumber \\  \cos\left(\frac{v_0m\pi t}{L}\right)\sin\left(\frac{v_0m\pi t}{L}\right) \nonumber
\end{align}
Taking into account that at initial condition $\bm{R}(t=0)=\bm{R_0}$ the water drop is a sphere with radius $R_0$ , then the final solution can be written as
\begin{equation}
\bm{R}=\bm{R}_0+\psi(\xi,t)\bm{S^\xi} \label{18}
%(\psi(x,t)\sin\phi \sin\theta, \psi(y,t)\sin\phi \cos\theta, \psi(z,t) \cos\theta) 
\end{equation}
%$0\leqslant\theta\leqslant2\pi, 0\leqslant\phi\leqslant\pi$ are spherical coordinates. 
The solution (\ref{18}) is not unexpected, but precisely explains oblate-prolate  oscillation of the drop observed in experiments. 

Since (\ref{13},\ref{14},\ref{18}) are particular solutions, then their linear combination is a general solution. Therefore, it predicts non-linear oscillations among spherical, oblate/prolate and non-oblate/prolate shapes, tendency towards increasing the radius, and inducing shape instabilities by forming singularities, and with combination of numerical solution to (\ref{6},\ref{normal}) \cite{grinfeld2010} indicates the growing amplitude in the oscillations of the surface mass density, meaning increasing the mass instabilities in diffusive layer of the surface. Mass instabilities in diffusive layer and linear combination of (\ref{13},\ref{14},\ref{18}) solutions ultimately lead to development of singularities in the drop, which may induce drop division.

Also, despite neglecting uniform translational motion of the drop (one may argue that for sufficiently small periods of time the inertial effects on drop shape motion is negligibly small), we might still speculate why oscillating surfaces can lunch water drop at higher speed, than hard sphere \cite{raufaste_2017}. Due to the generality of the DVS equations, an oscillating surface undergoes the same shape motion at the interface of substrate/droplet as water drops. If frequencies of the substrate surface and water drop matched, within an order of magnitude, then a resonance effect takes place and the drop will be launched by higher speed. We should also note that according to (\ref{4}-\ref{9}) and (\ref{10}) equations gravitational term $\rho gh$ has no effect on governing shape equations. Gravity has no role therefore surface motion would be exactly the same if one would remove gravity. This explains why water drops can move against gravity \cite{PhysRevLett.99.144501} on oscillating superhydrophobic substrate.   

We have proposed a system of nonlinear equations to describe the nonlinear features of the dynamics of two dimensional surfaces, with large deformations and large variations in density, both spatial and temporal \cite{svintradze_2016, svintradze_2017}. Analytic solutions to simplified version of two dimensional surface equations in gravity describe freely falling water drop's shape dynamics and precisely explain why the shape non-linearly oscillates among prolate/oblate and non-prolate/oblate forms and displays a wide range of shape instabilities. Surprisingly, despite the fact that the proposed system purposefully disregards the tangential components of surface velocities, the solution qualitatively explains experimentally observed shape dynamics \cite {deGennes2004, raufaste_2017, Chaudhury1539, PhysRevLett.94.068301, doi:10.1021/la046886s, PhysRevLett.102.194504, PhysRevLett.99.144501, Vollmer2014, richard2002}. By this analytic solution we have reproduced numerically solved Navier-Stokes equations results \cite{agrawal_2017}, but avoided  intensive modeling for surface pressure, which would be necessary for numerically solving Navier-Stokes. We have shown that the continuity equation is an approximation to the DVS equations, which, in contrast to Navier-Stokes, can be trivially handled in this concrete case. Also, we have shown that linear models leading to wave equations \cite{isenberg1992, chomaz1998} are generally true for near to equilibrium approximations.

The numerical solution of (\ref{6},\ref{normal}) combined with analytical ones (\ref{13},\ref{14},\ref{18}) directly imply that: surface mass density is non-linearly increasing while the shape adopts diverse forms. Impact of this statement in biology is that: if one starts formation of cell from mixture of organic molecules in a droplet and lets evolution of shape dynamics by applying some potential field (like hydrophobic-hydrophilic interactions for instance), then droplet will ultimately develop some membrane like structure (because surface mass density is nonlinearly increasing) and will adopt rich divers shapes. Such active dynamics may explain why cells got membranes \cite{Zwicker2016} and is in contradiction with generally thought idea that membranes were formed first. %even though prototype cell did not necessarily had one.  

\begin{acknowledgments}

We thank Dr. Julicher (MPIPKS), Dr. Chu (MPIPKS) and Dr. Roldan (MPIPKS) for discussions. This work was initiated at the Aspen Center for Physics, which is supported by National Science Foundation grant PHY-1607611 and was partially supported by a grant from the Simons Foundation.
 
\end{acknowledgments}

\bibliographystyle{apsrev4-1}
\bibliography{References}

\end{document}